# Sensitive singular-phase optical detection without phase measurements with Tamm plasmons

Svetlana V. Boriskina[*] and Yoichiro Tsurimaki

*Department of Mechanical Engineering, Massachusetts Institute of Technology, Cambridge, MA, 02139, USA*
*Corresponding author: sborisk@mit.edu*

**Abstract:** Spectrally-tailored interactions of light with material interfaces offer many exciting applications in sensing, photo-detection, and optical energy conversion. In particular, complete suppression of light reflectance at select frequencies accompanied by sharp phase variations in the reflected signal forms the basis for the development of ultra-sensitive singular-phase optical detection schemes such as Brewster and surface plasmon interferometry. However, both the Brewster effect and surface-plasmon-mediated absorption on planar interfaces are limited to one polarization of the incident light and oblique excitation angles, and may have limited bandwidth dictated by the material dielectric index and plasma frequency. To alleviate these limitations, we design narrow-band super-absorbers composed of plasmonic materials embedded into dielectric photonic nanostructures with topologically-protected interfacial Tamm plasmon states. These structures have planar geometry and do not require nanopatterning to achieve perfect absorption of both polarizations of the incident light in a wide range of incident angles, including the normal incidence. Their absorption lines are tunable across a very broad spectral range via engineering of the photon bandstructure of the dielectric photonic nanostructures to achieve reversal of the geometrical phase across the interface with the plasmonic absorber. We outline the design strategy to achieve perfect absorptance in Tamm structures with dissipative losses via conjugate impedance matching. We further demonstrate via modeling how these structures can be engineered to support sharp asymmetric amplitude resonances, which can be used to improve the sensitivity of optical sensors in the amplitude-only detection scheme that does not require use of bulky and expensive ellipsometry equipment.

## 1. Introduction

Bio(chemical) sensors with optical transduction often rely on measuring the frequency shifts of the optical modes caused either by the changes in the ambient refractive index or by the adsorption of molecules on the sensor surface [1,2]. Environmental changes to be detected are often very small, and require interactions of light with the target material over long distances in order to accumulate large enough phase change that translates into the measurable optical mode frequency shift. Large phase accumulation can be achieved over centimeters- or meters-long propagation lengths in optical fibers [3], while optical microcavities that trap and recycle light offer more compact micron-sized on-chip sensor designs [1,4,5]. The use of metal nanoparticles supporting surface plasmon-polariton modes enables further shrinking of the light-matter interaction volume, as strong localized field enhancement increases sensitivity of the particles spectral response to environmental changes [2,6,7]. The increase in sensitivity, however, is accompanied by the decrease in the spectral resolution, as surface plasmon modes dephase quickly due to high level of scattering and dissipative losses in metals and other plasmonic materials [8,9].

However, a phase of light is a cyclic variable, which is undefined at the point of complete destructive interference (a point of complete darkness), and varies rapidly in the vicinity of this point [10–12]. The vortex optical powerflow around phase discontinuities increases the light-matter interaction distance without increasing the device footprint [13,14]. Accordingly, light interference within a nanostructure that is accompanied by fast phase variations can be used to improve the sensitivity of bio(chemical) sensors with optical transduction [14–17].

In particular, the phase shift of the plane wave reflected from a planar interface exhibits singular behavior at frequencies where the surface reflectance vanishes. This property of the phase stems from the general principle of causality, which is reflected in the Kramers-Kronig relations [18] connecting the spectral reflectance from the surface $R(\omega)$ with the phase shift of the reflected wave $\varphi(\omega)$:

$$\varphi(\omega) = \frac{\omega}{\pi} \int_0^\infty \frac{\ln(R(\omega')/R(\omega))}{(\omega')^2 - \omega^2} d\omega' \tag{1}$$

The extinction of the light reflection from a dielectric surface characterized by a local refractive index $n$ illuminated by $p$-polarized light can be achieved at a specific angle of illumination known as the Brewster angle ($\tan(\theta_B) = n$). However, the Brewster extinction condition is not easily achievable with lossy materials ($n = n_r + in_i$), as it requires simultaneous cancellation of both the real and the imaginary parts of the reflection coefficient. The perfect absorptance condition can nevertheless be realized in thin-film metamaterial absorbers by tuning their effective refractive index, and is guaranteed by the Jordan topological theorem [15]. However, plasmonic metasurfaces with tailored optical properties enabling the perfect absorption condition often require complex electromagnetic design and low-throughput fabrication techniques such as electron beam lithography [15]. Plasmonic metamaterials prepared by the methods of colloidal chemistry and self-assembly [16,19,20] may still be restrictive in the spectral positioning of absorption resonances and require several fabrication steps to be realized.

It should be noted however that Eq. 1 is not limited to the absorbers characterized by an effective local refractive index, which significantly widens the design space for engineering the singular-phase reflection condition. Here, we demonstrate simple planar multilayered photonic-plasmonic structures that do not rely on the effective refractive index matching and instead are tuned to exhibit zero reflection due to the excitation of Tamm plasmon modes [21–32]. Tamm plasmons are surface states that can exist on the interface between the metal surface and a dielectric Bragg mirror (i.e., a one-dimensional photonic crystal). The existence of Tamm plasmon states is protected by the topological properties – manifested as geometrical Zak phases – of the optical bands of the photonic crystal [33–36] rather than by the Jordan's theorem as for the effective-index-matched thin films. The Tamm plasmon sensors can be easily designed to operate at a chosen frequency (or multiple frequencies), can be composed of a variety of plasmonic and dielectric materials, are amenable to fast and large-scale fabrication by either sputtering of vapor deposition techniques, and have high tolerance to fabrication imperfections [22,37].

Light localization and high local density of photon states in the Tamm plasmon modes have already been exploited in applications ranging from lasing [38] to fluorescent spectroscopy [39]. Furthermore, narrow reflection-minima features associated with the Tamm plasmon excitation have been used for refractive index sensing [29,40–42] and non-contact temperature-monitoring [22,43]. We have recently experimentally demonstrated the use of Tamm super-absorbers as singular-phase optical temperature sensors with over two orders of magnitude enhanced sensitivity over standard amplitude-measurement-based sensors [37]. However, these sensors exploited resonant complete absorptance condition for only one (*p*) polarization of light and required the use of ellipsometric equipment to perform phase-sensitive measurements. This is impractical for most end-user applications, where cheap and compact sensors relying on single-frequency amplitude-only measurements. Here, we demonstrate a new optical sensing scheme based on the amplitude measurements, which offers superior sensitivity to refractive index changes on the sensor surface.

## 2. Tamm plasmon properties: conjugate impedance matching yields phase singularity

Figure 1a shows a schematic of the Tamm structure comprised of a multi-layer dielectric Bragg reflector – a one-dimensional (1D) photonic crystal (PhC) – on top of a surface of bulk metal. The 1D PhC is composed of alternating low- and high-index thin dielectric films, marked as A and B layers, respectively. We simulate plane wave interactions with such planar multi-layer structures by using semi-analytical transfer matrix method [44]. The reflection coefficients of a multi-layered structure $R = |r_1|^2$ are calculated for both polarizations of the incident plane wave by using a recursive formula:

$$r_n = (\rho_n + r_{n+1}e^{-2ik_nd_n})/(1 + \rho_n r_{n+1}e^{-2ik_nd_n}), n = N, N-1, \dots, 1, \qquad (2)$$

where $\rho_n$ is the *p*- or *s*-polarized Fresnel reflection coefficient of the *n*-th material interface, $k_n$ is the normal component of the wavevector in the *n*-th medium, $d_n$ is the thickness of the *n*-th layer, and *N* is the total number of layers in the structure. The recursion is initialized by setting $r_{N+1} = \rho_{N+1}$. The calculated wavelength reflectance spectrum of the plane wave incident from the top normally to the surface of the Tamm structure ($\theta = 0^o$) is shown in Fig. 1b for the case of a 5-periods long dielectric PhC section made of 125nm-thick silicon dioxide ($SiO_2$) layers and 65nm-thick silicon carbide (SiC) layers on top of bulk gold (Au) substrate. The reflectance spectra from the PhC and Au surfaces separately are also plotted for comparison. The materials parameters for SiO$_2$ and SiC are taken from Palik [45]. Both dielectrics have negligible dissipation losses in the optical and near-infrared spectral ranges considered in this work, and their refractive indices vary slowly, averaging about *n*=1.45 for SiO$_2$ and *n*=2.6 for SiC. Experimental material dispersion properties of Au depend on the deposition process and metal layer thickness. To avoid accounting for the specifics of the deposition process, we calculated the dielectric function of Au via a Drude model $\varepsilon_{Au}(\omega) = \varepsilon_\infty - \omega_p^2(\omega^2 + i\gamma\omega)^{-1}$, with $\varepsilon_\infty = 9.84$, plasma frequency $\omega_p = 9eV$, and collision frequency $\gamma = 67meV$ [46]. The design approach outlined below is however completely general and imposes no limitations on either material properties or thicknesses of dielectric and metal layers.

The Tamm plasmon mode excitation in the structure shown in Fig. 1a is manifested by a narrow dip in its reflectance spectrum (Fig. 1b, red line), which appears in the frequency range overlapping with the photonic bandgap of the photonic crystal in the absence of the Au mirror (Fig. 1b, blue line). The choice of the operational wavelength is arbitrary, and the resonance can be easily tuned across the visible and infrared spectral ranges by adjusting the parameters of the structure (i.e., layer thicknesses or materials) [37]. Unlike surface plasmon polariton modes that are *p*-polarized, Tamm plasmon modes can be excited by incident waves of both orthogonal *s*- and *p*- polarizations [21,22]. At normal incidence of the plane wave, the reflectance minima for both polarizations coincide (Fig. 1c). The two resonant dips spectrally separate and move away from each other at larger angles of incidence, although they overlap again at the incident angle around 50º, before separating further at larger angles. The photonic band structure of the PhC stack without the Au mirror is shown in Fig. 1d, and reveals shrinkage of the PhC band gap for the *p*-polarized waves due to the Brewster effect at the interface between low and high index layers [47]. In turn, Fig. 1e shows the dispersion characteristics of the Tamm structure for both polarizations, and illustrates formation of Tamm interfacial states and their spectral shifts caused by the change of the angle of incidence. At small angles close to normal incidence, the Tamm states have parabolic dispersion with different 'effective masses' for *s*- and *p*-polarizations [21,22]. However, the *p*-polarized Tamm resonance blue shifts more significantly at larger angles of incidence than the *s*-polarized one due to the spectral shrinkage of the band gap for the *p*-polarized waves.

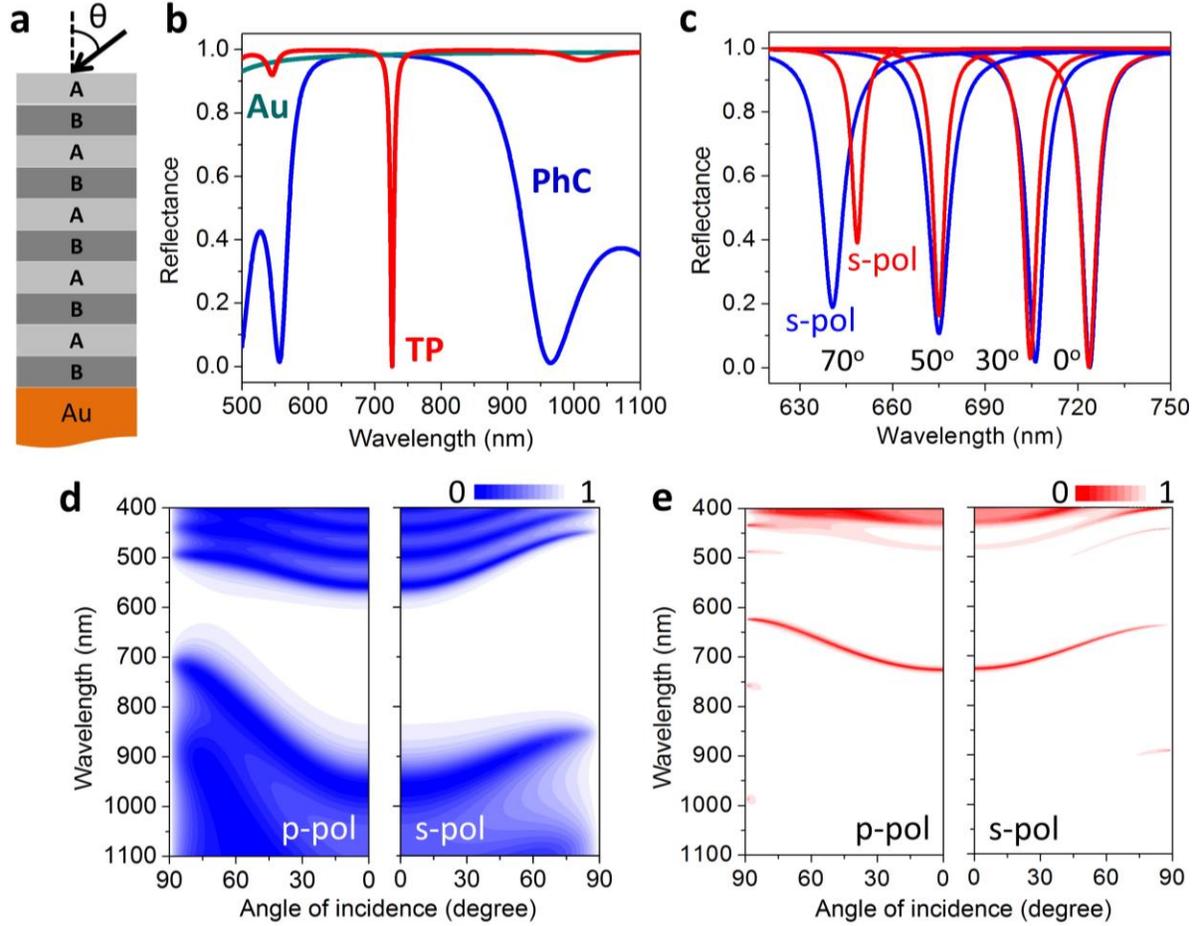

**Figure 1. Tamm plasmon resonances in hybrid metal-dielectric multilayer structures**. (a) A schematic of the Bragg stack of alternating dielectric layers of low (A) and high (B) refractive indices deposited on top of a metal surface. (b) Optical reflectance spectrum for the normal light incidence on the 5-period PhC section composed of 125nm-thick $SiO_2$ layers and 65nm-thick SiC layers on the Au substrate. The thickness of the SiC layer adjacent to the Au surface is tuned to 46.775nm to achieve the complete reflectance extinction. (c) Reflectance spectra of the Tamm structure for several angles of incidence (shown as labels) and for both polarizations. (d) Reflectance of the 5-period PhC section as a function of wavelength and angle of incidence, which exhibits photonic bandgaps for both polarizations of electromagnetic field (white areas in the plot indicate the photonic bandgap). (e) Reflectance of the Tamm structure shown in panel (a) exhibits narrow resonant features corresponding to the excitation of the Tamm plasmon states in the wavelength and angular range corresponding to the photonic bandgaps in (d).

Perfect absorption of incident light due to excitation of the Tamm state on the interface between the metal surface and the dielectric PhC can be achieved if the surface impedance of the bottom surface of the PhC facing the metal is tuned to conjugate-match that on the metal surface, $Z_{PhC} = Z^*_{metal}$ [34,37]. Conjugate impedance matching is a standard approach in the transmission lines theory, which is commonly used to maximize power transfer from the source to the load [48]. The same approach can also be applied to design nanoantennas and photon absorbers in the optical domain [49,50]. It should be noted that the optical surface impedance of a material interface is directly related to the material topological properties through the geometrical (Zak) phases of its bulk photonic bands [36]. Accordingly, it can be proven that interfacial states formed as a result of conjugate impedance matching at the interface are topologically protected and thus very robust to structure deformations [36,37].

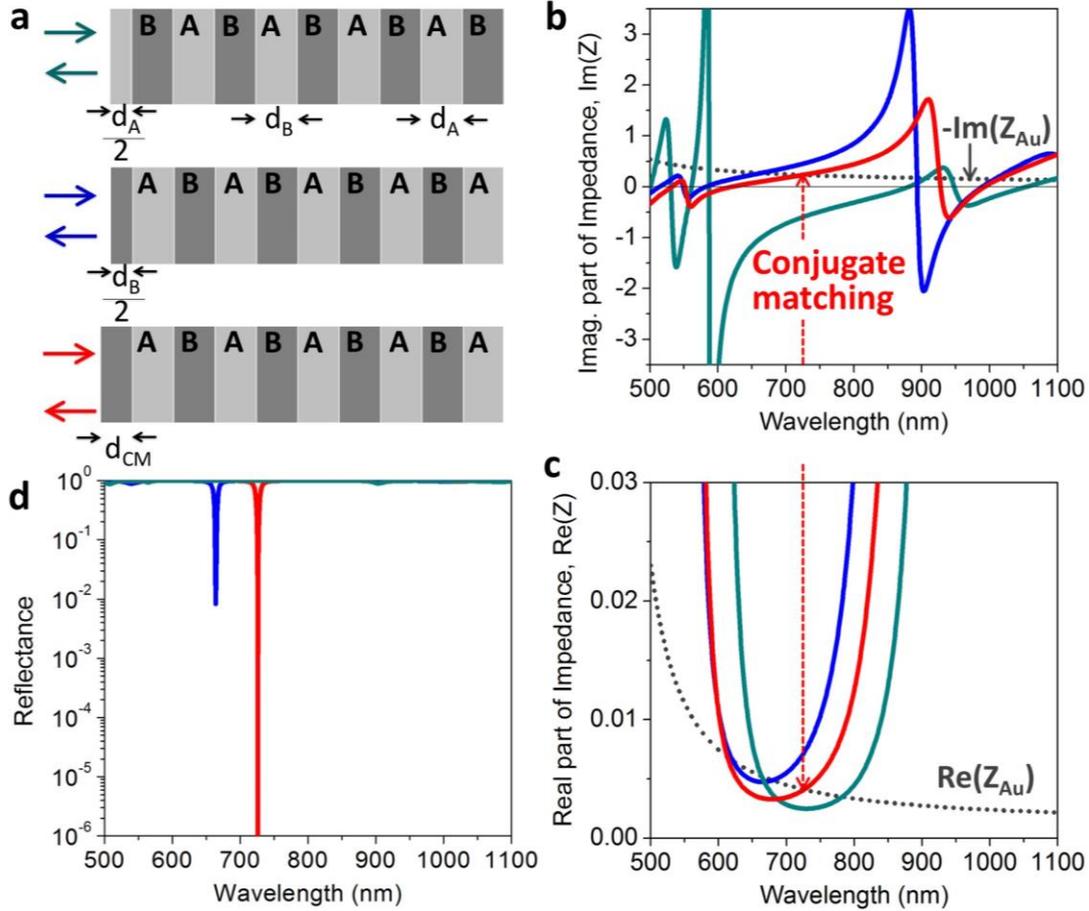

**Figure 2. Conjugate impedance matching creates an interfacial Tamm state.** (a) Various PhC termination configurations breaking the structure symmetry. (b,c) Frequency spectra of the real (c) and imaginary (b) parts of the complex optical impedance of the surface of bulk Au and the PhC surface with varying termination configurations shown in (a). (d) Reflectance spectra of the Tamm structure shown in Fig. 1a, which exhibits complete reflectance suppression only at the frequency of perfect conjugate-impedance matching between the two surfaces (i.e., when $Z_{PhC} = Z_{Au}^*$).

We have recently demonstrated that perfect Tamm absorbers can be designed by conjugate impedance matching finite-size 1D photonic crystals and metal surfaces [37]. Below their plasma frequency, metal surfaces are characterized by predominantly reactive impedances (i.e., with $Im\{Z_{metal}\} > Re\{Z_{metal}\}$). Metal surface impedance is strongly mismatched with the real-valued impedance of the free space (i.e., with $Im\{Z_{fs}\} = 0$), which is the reason metals strongly reflect incident light. Conjugate impedance matching of the metal surface to the PhC transforms the reflective surface into a resonant perfect absorber, whose top dielectric surface is automatically impedance-matched to the free space. At the wavelength where the perfect absorption condition is satisfied, the reflection phase becomes singular (see Eq. 1), while the reflection amplitude vanishes. This makes the optical response of the conjugate-matched Tamm structure very sensitive to any changes in its environment. Below, we provide a detailed discussion of the conjugate impedance matching strategy applied to guarantee and protect the Tamm mode at the PhC-metal interface.

Figure 2 illustrates how a 1D photonic crystal geometry can be tuned to achieve conjugate impedance matching with a gold interface, which yields formation of and efficient light coupling to the interfacial state shown in Fig. 1b. We would like to emphasize that formation of the interfacial Tamm plasmon states does not require periodicity in the medium, and can be achieved in completely aperiodic structures [22,37]. A 1D periodic PhC was chosen here to illustrate the conjugate-matching mechanism of the Tamm state formation owing to its well-studied geometry and known spectral characteristics. The thicknesses of the high and low index layers roughly measure a quarter of the wavelength in the corresponding media at center of the PhC bandgap, and were tuned to improve impedance matching. The PhC has been truncated to a short section of 5 periods, followed by optimizing the PhC termination point. Truncation of the PhC length is necessary to match the real part of the Au surface impedance, which is small but not negligible for real metals. In contrast, the surface impedance of a perfect loss-less infinite PhC is purely imaginary at frequencies within its stop bands [36]. Three positions of the termination point are shown in Fig. 2a, (i) in the center of the low-index dielectric layer (top panel), (ii) in the center of the high-index layer (middle panel), and (iii) at an optimized position within the high-index layer (bottom panel). Figures 2b,c depict evolution of the real and imaginary parts of the PhC surface impedance inside and around the photonic bandgap spanning from about 600nm to 900nm in wavelength. The complex impedance values of the gold surface are also shown for comparison. It can be seen that in the case of the optimized truncation point, the conjugate impedance matching condition (Figs. 2b,c, red line) leads to the complete extinction of the reflectance from the interface, owing to the optical energy coupling into the interfacial state (Fig. 2d). In can also be seen that by matching only the imaginary part of the impedance, coupling – albeit not ideal – to the interfacial state can also be achieved (Figs. 2b-d, blue line). However, matching of only the real parts of the impedances does not result in the surface state formation (Figs. 2b-d, teal line).

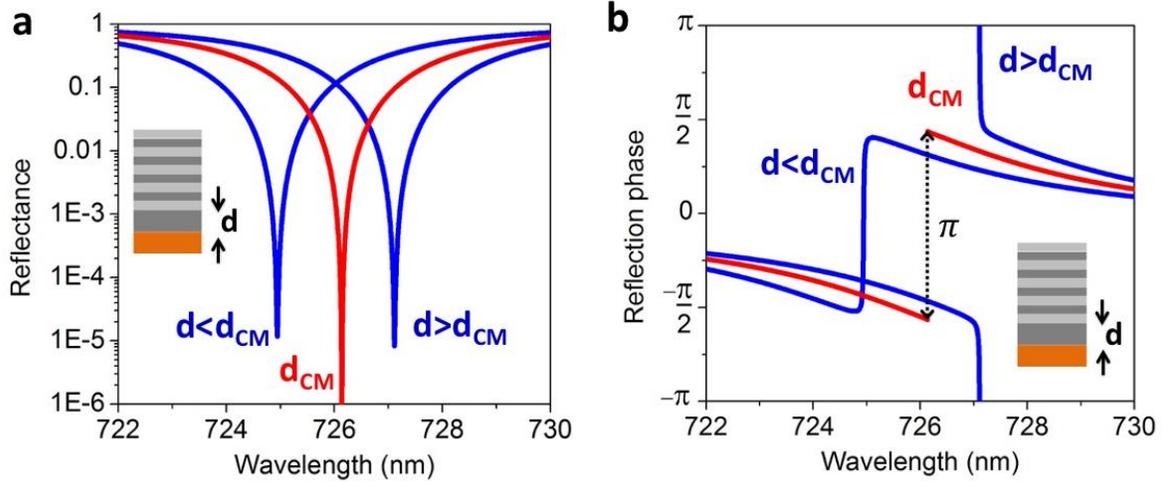

**Figure 3. Zero-reflection and singular-phase response of the impedance-matched Tamm structure.** (a) Reflectance of the structure shown in Fig. 1a in the frequency range around the Tamm mode resonance for the optimum termination point yielding conjugate-matched structure ($d=d_{CM}=46.775$nm, red line) and two near-optimum points ($d=46.5$nm and $47$nm, blue lines). (b) The reflection phase spectrum corresponding to the conjugate-matching structure (red line) and near-matching ones (blue lines).

Figure 3b illustrates the changes in the reflection phase that accompany the zero- and near-zero reflectance condition. The structure is tuned across the optimum matching point by slightly varying the thickness $d$ of the high-index termination layer. Phase discontinuity can be clearly observed if the zero-reflectance condition is satisfied at the optimum termination point ($d = d_{CM}$, red lines), when the reflectance amplitude vanishes (see also Eq. 1). It is of course impossible to control the thickness of the layer with such precision during the fabrication process, but it also is not necessary. Owing to the topological protection of the Tamm state, even significant deviations of the structure geometry from the optimum configuration cannot destroy the surface mode resonance. This is evident in Fig. 3a and Fig. 2d, where the sharp deep resonances are still observed in the both the blue curves, which correspond to the reflection spectra of two non-perfect Tamm structures (see also Fig. 2d). It should be noted that the spectral position of the resonance can be further tuned post-fabrication by adjustments of the angle of incidence [22,37]. Even though singular-phase point cannot be achieved under experimental conditions, operating close to this point still allows making use of the sharp phase variations (both the blue lines) to increase Tamm sensor sensitivity.

## 4. Singular-phase optical sensing with and without phase measurements

The sharp variations in the reflection phase observed in Fig. 3b provide a useful transduction mechanism to achieve optical sensing with higher sensitivity than that of the conventional amplitude-based sensing scheme [15,37,51]. Prior experiments on the singular-phase optical sensing relied on the use of ellipsometers to detect not only the amplitudes $|r_{p,s}|$ but also the phases $\delta_{p,s}$ of the reflected waves of both orthogonal *s* and *p* polarizations ($r_{p,s} = |r_{p,s}|e^{i\delta_{p,s}}$) by measuring two ellipsometric angles $\Psi = \mathrm{atan}(|r_p|/|r_s|)$ and $\Delta = \delta_p - \delta_s$. As such, these experiments typically used an oblique-angle excitation configuration [15,37]. This is a typical excitation configuration in the surface-plasmon spectroscopy and Brewster interferometry [15,52–54], due to the large SPP momentum that needs to be matched to achieve coupling, and the large values of the Brewster angles of common materials and meta-materials, respectively.

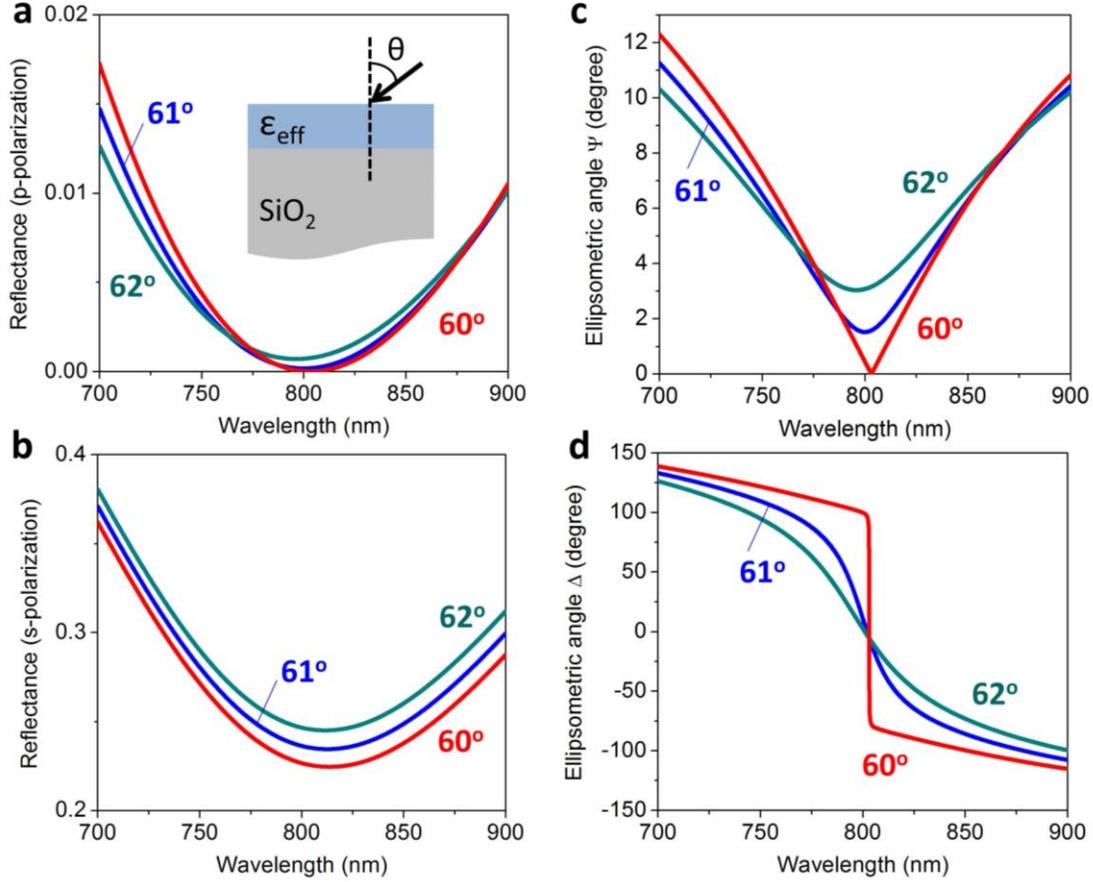

**Figure 4. Singular-phase reflectance and ellipsometric angles spectra of the Brewster-type sensor based on an effective-index-engineered thin-film absorber.** (a,b) Reflectance spectra for the *p*-polarized (a) and *s*-polarized (b) plane waves incident on a structure shown in the inset of panel a at the Brewster angle of 60 degrees and near-Brewster angles of 61 and 62 degrees (shown as labels). The film thickness is 170 nm and its complex effective permittivity is 6.37+*i*0.5. (c,d) Amplitude (c) and phase (d) ellipsometric angles corresponding to the reflectance spectra shown in panels a,b.

Typical reflectance and ellipsometric angles spectra of a Brewster-type singular-phase sensor based on a thin film with a complex effective refractive index are shown in Fig. 4. The thin film can be comprised of either nano-patterned plasmonic antennas or of metal nanoparticles dispersed in a dielectric matrix. The singular-phase operation is achieved at a large oblique Brewster angle (60 degrees to normal), where the *p*-polarized plane wave exhibits zero reflection (Fig. 4a). At the frequency of zero reflection, the amplitude ellipsometric angle also reduces to zero, while the phase ellipsometric angle exhibits a singular behavior (Figs. 4c,d) useful for the singular-phase optical sensing [15]. However, the sharp phase variation observed at the Brewster angle (red line in Fig. 4d) quickly disappears at the illumination angles deviating from the Brewster angle even slightly. Likewise, Brewster angles of natural materials are highly oblique and uniquely determined by the material properties, making near-normal angle sensor excitation impossible. Our prior experiments on singular-phase sensing with Tamm structures relied on the use of ellipsometers to measure the reflectance phase [37]. Accordingly, the Tamm structures have been optimized to exhibit zero-reflection resonances at large oblique incident angles suitable for the ellipsometric measurements. Under such condition, the Tamm structure exhibits complete zero-reflectance singular-phase resonance for only one polarization, with the amplitude ellipsometric angle spectrum similar to that shown in Fig. 4c.

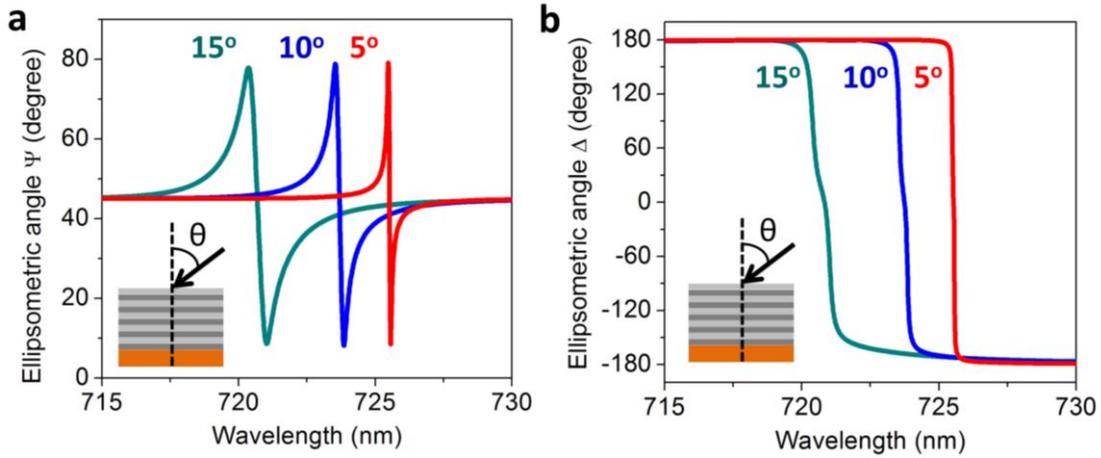

**Figure 5. Ellipsometric angles spectra of the Tamm structure at near-normal incident angles.**
(a,b) Amplitude (a) and phase (b) ellipsometric angles of the structure from Fig. 1 for several close-to-normal angles of incidence shown as labels.

Here, we propose to exploit the fact that – unlike the Brewster angle reflectance condition – zero-reflectance and singular-phase conditions for the Tamm plasmons can be satisfied for both polarizations of the incident light (Fig. 1c). At incident angles closer to normal, the frequencies of the *s*- and *p*-polarized interfacial states are closer to each other, and both states can be efficiently excited yielding two resonant wavelengths exhibiting near-zero reflectance from the structure. As a result, not only the phase ellipsometric angle but also the amplitude ellipsometric angle exhibits the asymmetric line shape. These line shapes are plotted in Fig. 5 for the structure shown in Fig. 1 and several excitation angles close to normal. At all the angles, a sharp asymmetric spectral feature is observed in Fig. 5a, which becomes sharper as the normal incidence condition is approached. Note that the variations in the phase ellipsometric angle spectra in Fig. 5b span the angular range of $2\pi$ (as compared to $\pi$ for the Brewster-angle resonances in Fig. 4d), while the rapid variations in the amplitude ellipsometric angle spectra span the angular range approaching that from 0 to $\pi$ (Fig. 5a). The linewidth of this asymmetric spectral feature increases with the increasing angle of incidence, and the deviation of the resonance condition at non-normal incidence from that of the complete absorptance reduces the angular range to the value just under the 0 to $\pi$ interval.

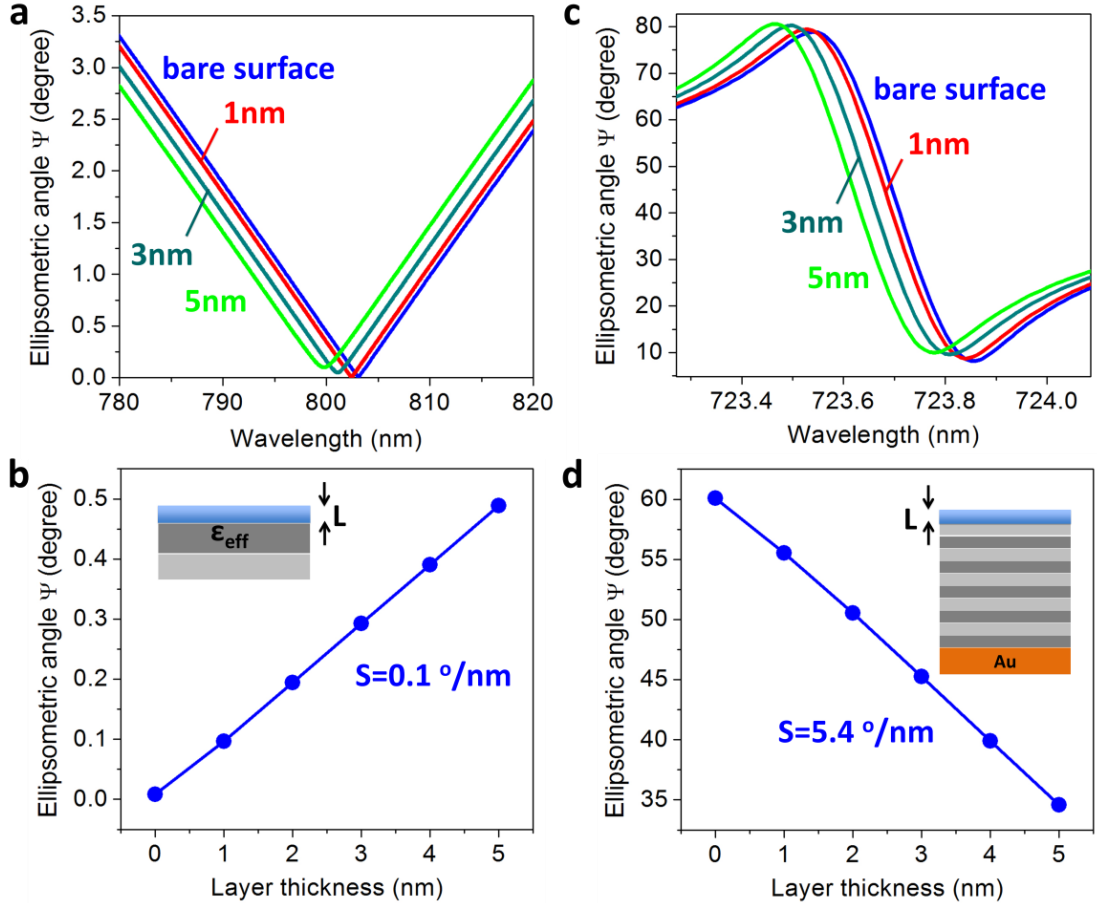

**Figure 6. Sensitive optical detection with Tamm absorber without phase measurements.** (a) Amplitude ellipsometric angle spectra of the Brewster-type sensor shown in the inset of Fig. 4a for the incident angle of 60 degrees to normal. Different color spectra correspond to the bare Brewster sensor (teal line), and the structure with a thin layer of low-index material ($n=1.5$) adsorbed on the sensor top surface (blue and red lines). (b) Ellipsometric angle change with the increase of the adsorbed layer thickness. (c) Amplitude ellipsometric angle spectra of the Tamm structure for the incident angle of 10 degrees to normal. (d) The change of the ellipsometric angle with the increase of the thickness the dielectric layer formed on top of the Tamm structure at wavelength of 723.65nm. The insets to panels b and d show the schematics of both sensors with the low-index layer on top.

Amplitude ellipsometric angle $\Psi = \operatorname{atan}(|r_p|/|r_s|)$ can be acquired experimentally without phase-sensitive measurements, by simply measuring the reflectance for both polarization of the incident light at the resonant wavelength corresponding to the Tamm state excitation. However, the use of ellipsometer is necessary if the phase measurements are also of interest. In Fig. 6, we demonstrate how the sharp asymmetric resonances in the amplitude ellipsometric angle can be used to achieve sensitive optical detection of the ultra-thin layers of molecules adsorbed to the Tamm structure surface [1,55]. We also compare the performance of the Tamm sensor with the Brewster-type sensor based on the structure shown in Fig. 4. We assume a low refractive index of $n = 1.5$ for this adsorbed dielectric layer, which is a typical value for many organic biological materials [1,56]. Figure 6a shows several wavelength spectra of the amplitude ellipsometric angle for varying thicknesses of the low-index dielectric layer on top of the Brewster sensor with the same parameters as in Fig. 4, and light illumination at 60 degrees to normal. The spectrum of the bare Brewster interferometer without the dielectric layer is also shown for comparison as the blue line. Since the zero-reflectance minimum is observed in the Brewster-type sensor only for the $p$-polarized light, the amplitude ellipsometric angle in Fig. 6a exhibits a typical dip in its spectrum. The depth and the spectral position of this dip changes with the increased thickness of the adsorbed layer (Figs. 6a,b). However, these changes are very small, yielding low sensor sensitivity $S_L = \Delta\Psi/\Delta L = 0.1°/nm$.

In contrast, the corresponding changes in the asymmetric spectral feature in the amplitude ellipsometric angle of the Tamm structure are on the scale of the tens of degrees (Fig. 6c). The variation of the Tamm structure amplitude angle is plotted in Fig. 6d as a function of the layer thickness, and shows the high sensitivity ($S_L = 5.4°/nm$) of the proposed amplitude-only detection scheme. Comparing Fig. 6b and d we can observe an over 50-fold increase in the sensitivity of the Tamm sensor tuned to operate at near-normal light incidence over the conventional Brewster-type interferometric sensor operating under oblique illumination at large angles.

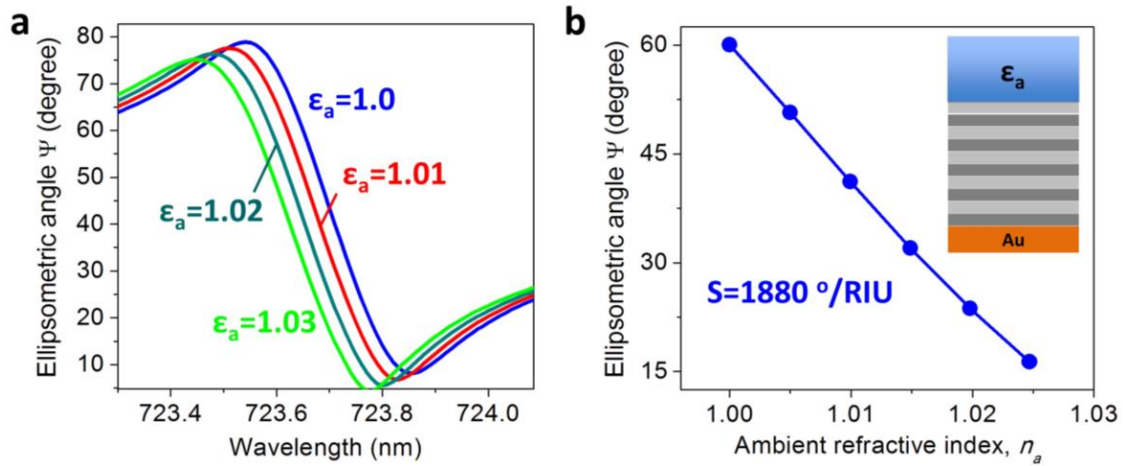

**Figure 7. Refractometric detection with Tamm sensor without phase measurements.** (a) Amplitude ellipsometric angle spectra of the Tamm structure for the incident angle of 10 degrees to normal and varying dielectric constant of the ambient medium. (d) The change of the ellipsometric angle with the increase of the ambient refractive index at wavelength of 723.65nm. The inset shows the schematic of a sensor with the varying-index ambient medium filling the upper half-space.

Figure 7 further illustrates that the same sensor can be used to detect changes in the ambient refractive index $n_a$ (or dielectric constant, $\varepsilon_a = n_a^2$) rather than the adsorption of molecules on the sensor surface. In this scenario, the sensitivity is defined as $S_n = \Delta\Psi/\Delta n_a$, and is estimated to reach $S_n = 1880^o/RIU$ at the excitation angle of 10 degrees to normal.

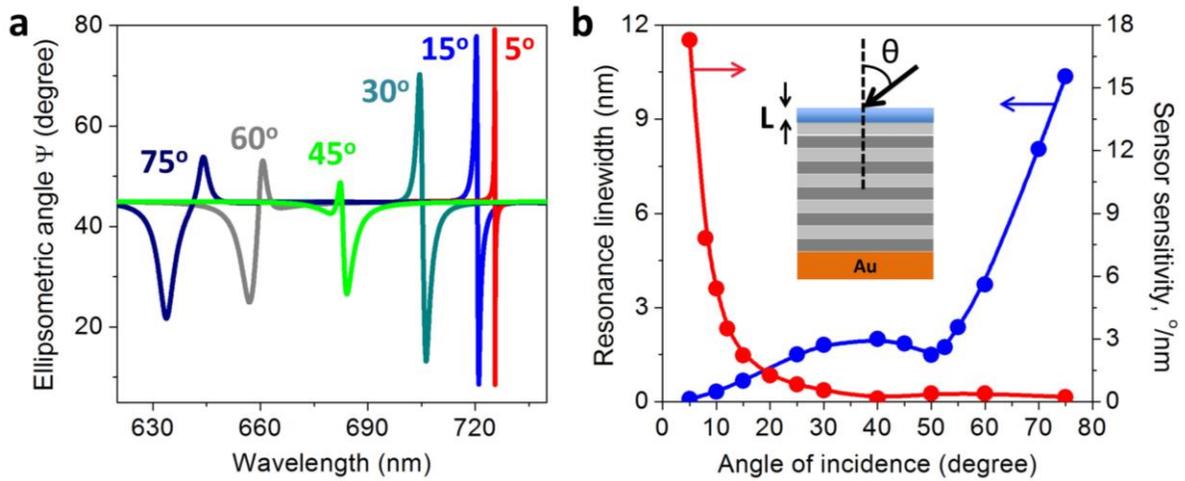

**Figure 8. Sensor sensitivity dependence on the optical excitation angle.** (a) Amplitude ellipsometric angle spectra of the Tamm structure for several values of the incident angle of light, shown as labels on individual spectral curves. (b) Evolution of the linewidth of the resonant asymmetric spectral feature as a function of the angle of incidence (blue line, left vertical axis) and the corresponding sensor sensitivity to the adsorbed layer thickness (red line, right vertical axis). The inset shows the schematic of the sensor and the incident angle definition.

Figure 8 illustrates the dependence of the asymmetric resonance linewidth and the Tamm sensor sensitivity to the angle of incidence, including highly oblique angles typical for the Brewster-type sensors. The 'linewidth' of the asymmetrical resonant feature is defined as the wavelength difference between the resonant peak and the closest neighboring dip [57,58]. It can be seen that as the angle deviates from the normal, the asymmetric spectral feature becomes broader. Then, at an angle of about 50 degrees, the Tamm dispersion branches of the two opposite polarizations cross before separating again at larger oblique angles. In turn, the sensor sensitivity maximizes as the angle approaches that of normal incidence, and reduces dramatically at oblique angles. Note that if the sensor is optimized to exhibit complete absorption for one polarization of incident light at a larger angle of incidence [37], the resonant spectral feature in the amplitude ellipsometric angle spectrum becomes symmetrical, and the amplitude-detection sensitivity is reduced.

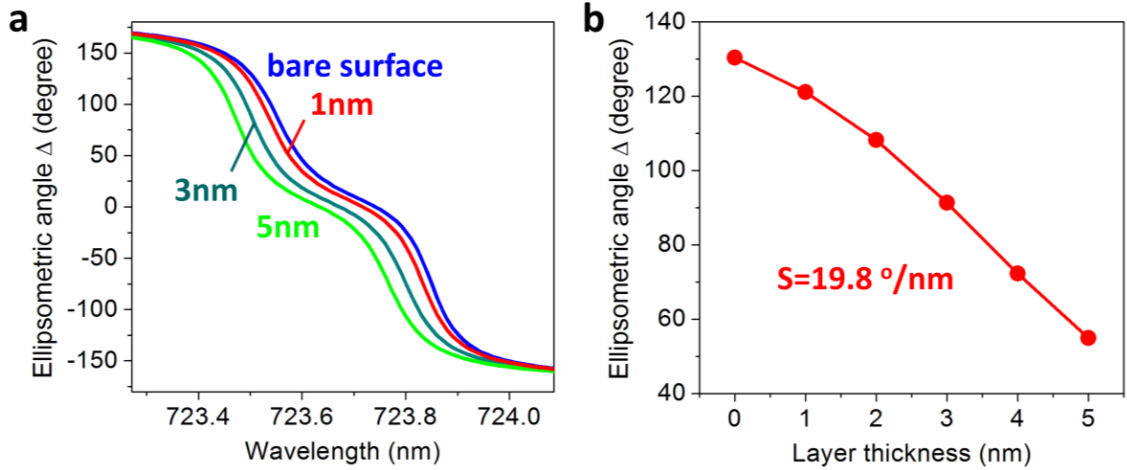

**Figure 9. Near-singular phase detection with Tamm structures.** (a) Phase ellipsometric angle spectra of the Tamm structure for the incident angle of 10 degrees to normal. (b) The change of the phase ellipsometric angle with the increase of the thickness the dielectric layer formed on top of the Tamm structure at wavelength of 723.5nm.

It should be noted that phase measurements could further increase the sensitivity of the Tamm sensor, as illustrated in Fig. 9, which shows the change of the phase ellipsometric angle with the increase of the adsorbed layer thickness. The estimated sensitivity increase is about four-fold if the phase measurement scheme is used. In comparison, Tamm sensors operated at large incidence angles exhibit over an order of magnitude drop in the sensor sensitivity between the phase and amplitude detection schemes [37]. We can conclude that although the phase measurement scheme yields higher sensitivity under both sensing scenarios, the structures optimized to exhibit two near-overlapping resonances enable amplitude-only measurements with enhanced sensitivity. This effect is general, and not limited to the Tamm sensor design in particular. However, the simplicity of the Tamm structures makes them attractive candidates to realize high-sensitivity amplitude-only sensors with asymmetric resonant line shapes.

**Conclusions**

We demonstrated through rigorous modeling how sharp phase and amplitude changes at the Tamm plasmon resonant frequency can be exploited to achieve sensitive optical detection. Tamm plasmon structures offer simplicity of the design, significant simplification of the fabrication processes over previously demonstrated singular-phase plasmonic metamaterial detection platforms, and superior tunability of the spectral position of the singular phase formation. The hybrid planar Tamm structures can be engineered to exhibit perfect absorption condition across a wide range of wavelengths, materials, layer thicknesses, and angles of incidence of the probe light [21,25,59]. This can be achieved by simple tuning of a few geometrical parameters such as the number of PhC periods, the PhC termination point, and the thickness of the plasmonic material layer. Since Tamm plasmon states can be excited by both *p*- and *s*-polarized optical fields, the amplitude-only detection scheme offers significant sensitivity improvements over the standard ones based on the SPP modes excitation or Brewster-angle sensing. Our calculations predict sensitivity of the Tamm detector of $S = 5.4°/nm$ in the near-infrared part of the optical spectrum and at light incidence at 10 degrees to normal, which is an over 50-fold enhancement over the Brewster-type sensor. Further improvements to the sensitivity are possible if the incident angle is even closer to the normal direction. Finally, we would like to emphasize that the proposed sensitivity-enhancement and conjugate-impedance matching mechanisms are very general and not limited to the Tamm structures considered in this paper. As such, they may lead to further research and new device demonstrations beyond the planar structures and refractometric sensors, which are the focus of this work.

## Acknowledgements

This work was supported by DOE BES Grant No. DE-FG02-02ER45977.